\documentclass[british,a4paper,10pt]{article}
\usepackage{amssymb}
\usepackage{amsmath}
\usepackage{subfigure}
\usepackage{slashed}
\usepackage{amsmath}
\usepackage[normalem]{ulem}
\usepackage{graphicx,psfrag}
\usepackage{mathrsfs}
\usepackage{amsfonts}
\usepackage{multirow}
\usepackage{hyperref}
\usepackage{placeins}
\usepackage{amsthm}
\usepackage{latexsym}
\usepackage[dvips]{epsfig}
\usepackage{enumitem}  
\usepackage{soul}
\usepackage{graphicx}
\usepackage{amsthm}
\usepackage{latexsym}
\usepackage[dvips]{epsfig}
\usepackage[utf8]{inputenc}
\usepackage{babel}
\usepackage{multicol}
\usepackage{color}
\usepackage{comment}
\usepackage[dvipsnames]{xcolor}
\usepackage{wasysym}
\usepackage{mathrsfs}
\usepackage{bm}
\usepackage{authblk}
\usepackage{slashed}
\usepackage{yhmath} 
\usepackage{stmaryrd}

\usepackage{pifont}
\usepackage{mdframed}
\theoremstyle{plain}
\newtheorem{proposition}{Proposition}
\newtheorem{lemma}{Lemma}

\newtheorem{assumption}{Assumption}

\newtheorem{corollary}{Corollary}

\newtheorem{remark}{Remark}
\def\bmeta{{\bm \eta}}
\def\bmg{{\bm g}}
\def\bmsigma{{\bm \sigma}}

\def\uL{\underline{L}}
\def\bme{{\bm e}}

\def\bmeta{{\bm \eta}}
\def\bmg{{\bm g}}
\def\bmsigma{{\bm \sigma}}

\def\ue{\underline{e}}
\def\uL{\underline{L}}
\def\bme{{\bm e}}
\def\bmue{\underline{\bme}}
\def\bmA{{\bm A}}

\allowdisplaybreaks
\begin{document}

\title{\textbf{Polyhomogeneous spin-0 fields in Minkowski spacetime}}
\author[1]{Edgar Gasper\'in   
  \footnote{E-mail address:{\tt edgar.gasperin@tecnico.ulisboa.pt}}}
  \affil[1] { CENTRA, Departamento de F\'isica, Instituto Superior
    T\'ecnico IST, Universidade de Lisboa UL, Avenida Rovisco Pais 1,
    1049 Lisboa, Portugal  }

\maketitle

\begin{abstract}
  The asymptotic behaviour of massless spin-0 fields close to spatial
  and null infinity in Minkowski spacetime is studied by means of
   Friedrich's cylinder at spatial infinity. The results are applied
  to a system of equations called the good-bad-ugly which serves
  as a model for the Einstein field equations in generalised harmonic
  gauge.  The relation between the logarithmic terms (polyhomogeneity)
  appearing in the solution obtained using conformal methods and those
  obtained by means of a heuristic method based on H\"ormander's
  asymptotic system is discussed. This review article is based
  on Class. Quantum Grav. 40 055002 (\cite{DuaFenGasHil22b})
  and arXiv:2304.11950  (\cite{GasPin23}).
\end{abstract}

\section{Introduction}

Asymptotic properties of the gravitational field encoded through the
geometric notion of null infinity is an ubiquitous element in several
open problems in general relativity. Despite that the notion of null
infinity was introduced in general relativity since the 1960's ---see
\cite{Pen63}--- the study of global properties and asymptotic
behaviour of the gravitational field is far from being a closed
subject \cite{Fri06, Val16}.  Moreover, since gravitational radiation
is only unambiguously defined at null infinity, the detection of
gravitational waves ---see \cite{AbbAbbAbb16}--- gives further
astrophysical motivation to the study of asymptotics and the global
structure of spacetimes.
One of the difficulties for the analysis of the global structure of
dynamical spacetimes (describing astrophysical objects) is that null
infinity is located infinitely far away from the gravitating sources
(say, a binary system) and hence evaluating the gravitational
radiation at $\mathscr{I}$ is non-trivial. However,
this does not mean that global analysis in general relativity (even
without resorting to conformal compactifications) is impossible as
clearly evidenced in prominent works on the global non-linear
stability of spacetimes of \cite{ChrKla93, DafHolRod21, LinRod04,
  HinVas18, HinVas20}.

\medskip

Nevertheless, the central idea behind the original notion of null
infinity ---as introduced in \cite{Pen63}--- is that of conformal
compactification. In such approach, properties of the physical
spacetime $(\tilde{\mathcal{M}},\tilde{\bmg})$ satisfying the Einstein
field equations are studied through an auxiliary ``unphysical''
spacetime $(\mathcal{M},\bmg)$ with $\bmg=\Xi^2\tilde{\bmg}$ so that
null infinity, denoted by the symbol $\mathscr{I}$, corresponds to the
region in $\mathcal{M}$ where $\Xi=0$ but $\mathbf{d} \Xi
\neq 0$. The latter condition ensures that, in asymptotically flat
spacetimes, $\mathscr{I}$ is composed by two disjoint hypersurfaces
called future and past null infinity ($\mathscr{I}^{\pm}$) with normal
covector $\mathbf{d} \Xi$. The points in $\mathcal{M}$ where
$\mathbf{d} \Xi$ vanishes are distinguished regions of the
\emph{conformal boundary} ($\Xi=0$) and correspond, in asymptotically
flat spacetimes, to spatial infinity ($i^0$), and
future and past timelike infinities ($i^\pm$).
A formulation of the
Einstein field equations that incorporates Penrose's proposal within
the initial value problem are Friedrich's conformal Einstein field
equations (CEFEs) ---see \cite{Fri81, Fri83, Fri02CEE, Val16}. This
formulation has been exploited in several applications in mathematical
relativity ---for instance \cite{Fri86, Fri83, Fri98a, Val04,
  ZhaHilVal21, GasVal18, LueVal13} to cite a few.  The CEFEs have also
been employed for numerical relativity evolutions in \cite{Hub99,
  Hub01, DouFra16, BeyFraSte17, FraSte21, FraSte23, FraGooSte23}.
Expanding on the latter point, it should be mentioned that, currently,
most numerical evolutions routinely used for astrophysical
applications do not make use of the CEFEs and, hence,
the conformal boundary is not included. This is because, at first
glance, the CEFEs look strikingly different to most standard
formulations of general relativity used for numerical applications.
However, there are different proposals to provide with alternatives to
the CEFEs that are closer to traditional formulations used routinely
in numerical relativity such as \cite{VanHus14, Zen08, Van15,
  HilHarBug16, GasHil18, GasGauHil19, DuaFenGasHil22a, MonRin08,
  BarSarBuc11} but which insist on including (a portion of) null
infinity.  One of them is the hyperboloidal-dual-foliation (HypDF)
approach of \cite{HilHarBug16}. One of the key requirements for such
formulation to work is that certain derivatives of a quantity called
the coordinate lightspeed (constructed directly from the metric),
decays fast enough towards null infinity.  To understand this point, in
\cite{GasHil18} a heuristic asymptotic system method was employed to
asses under which circumstances one can expect the coordinate
lightspeed condition to be satisfied. The notion of asymptotic system
is related to the weak null condition which played an important role
in the proof of the global non-linear stability of the Minkowski
spacetime in harmonic gauge \cite{LinRod04}.  The asymptotic system
concept has been key in the development of the HypDF program
\cite{HilHarBug16, GasHil18, GasGauHil19, DuaFenGasHil22a}.  In
particular, in \cite{DuaFenGasHil21} the asymptotic system method was
generalised to obtain asymptotic expansions for the gravitational field
and a polyhomogeneous peeling (i.e. presence of logarithmic terms in
the expansion for the Weyl scalars) result was derived in
\cite{DuaFenGas22}.

\medskip

Interestingly, the failure of the classical peeling result obtained in
\cite{DuaFenGas22} looks similar to that of \cite{GasVal18} which uses
a contrastingly different formal expansion scheme based on the CEFEs
and conformal methods.  To clarify the relation between these two
results, in \cite{DuaFenGasHil22b} the conformal method was applied to
a model system of equations which is called the good-bad-ugly (GBU),
upon which the bare-bones regularisation of the HypDF approach is
based ---see \cite{DuaFenGasHil22a}.  The basic conclusion from
\cite{DuaFenGasHil22b} is that, in the simplest case of the GBU-system
in Minkowski spacetime, the logarithmic terms in \cite{DuaFenGasHil21}
(and hence those of \cite{DuaFenGas22}) do not correspond to those of
\cite{GasVal18}. Nonetheless, to alleviate this shortcoming, a method
to retrieve the leading order logarithmic term within the asymptotic
system heuristics laid out in \cite{DuaFenGasHil22b}. Later, a method
to recover all the missed higher-order logarithmic terms (in flat
spacetime) was obtained as a subproduct of
the analysis of \cite{GasPin23} for the calculation of the
Newman-Penrose constants of spin-0 fields.  In this review article,
the discussion of the GBU-system close to spatial and null infinity of
\cite{DuaFenGasHil22b} and the higher order asymptotic system approach
trough conservation laws in flat spacetime discussed in
\cite{GasPin23} are revisited and presented in a coherent and
contiguous way.

\section{The good-bad-ugly model}

As briefly described in the introduction, the regularisation of the
HypDF equations in \cite{DuaFenGasHil22a} makes use of the
asymptotic system heuristics to predict the decay of the gravitational
field towards null infinity. In turn, the asymptotic system is related
to the concept of the weak null condition. The weak null condition,
introduced in \cite{LinRod03}, is a condition on the structure of the
non-linearities present in certain systems of wave equations for which
small data global existence has been proven.  The most important
example is that of the proof of the global non-linear stability of the
Minkowski spacetime in harmonic gauge of \cite{LinRod04}. Although, at
the time of writing, there is no proof (unlike the case for the classical null
condition) that the weak null condition is sufficient to establish
small data global existence results, this has been proven for a more
restricted version called the hierarchical weak null condition in
\cite{Kei17}. Hence, the expectation is that the satisfaction of the
weak null condition is at least indicative that obtaining a small data
global existence result should be possible.  Taking this as
motivation, in \cite{GasHil18} and then further developed in
\cite{DuaFenGasHil21}, the asymptotic system heuristics were employed
to devise a strategy for modifying the Einstein field equations in
generalised harmonic gauge by adding multiples of the constraints and
adapting the gauge so that the coordinate lightspeed condition of
\cite{HilHarBug16} is satisfied. The basic strategy of
\cite{GasHil18, DuaFenGasHil21} was built upon studying the following
model: consider three fields $\tilde{\phi}_{g}$,
$\tilde{\phi}_{b}$ and $\tilde{\phi}_{u}$, propagating in (physical)
Minkowski spacetime~$(\tilde{\mathcal{M}},\tilde{\bmeta})$ according
to 
\begin{subequations}\label{gbu}
\begin{align}
  & \tilde{\square} \tilde{\phi}_{g} = 0, \label{gbu-good}
  \\ &\tilde{\square} \tilde{\phi}_b = (\tilde{\nabla}_{\tilde{t}}
  \tilde{\phi}_g)^2 , \label{gbu-bad} \\ &\tilde{\square} \tilde{\phi}_u =
  \frac{2}{\tilde{\rho}}\tilde{\nabla}_{\tilde{t}} \tilde{\phi}_u. \label{gbu-ugly}
\end{align}
\end{subequations}
Here $\tilde{\nabla}$ is the Levi-Civita connection of
$\tilde{\bmeta}$ and $\tilde{\square}
:=\tilde{\eta}^{ab}\tilde{\nabla}_a\tilde{\nabla}_b$.  The fields with
labels ${}_{g}$, ${}_{b}$ and ${}_{u}$, will be referred to as the
~{\it good, bad} and {\it ugly} fields, respectively.  Since the
conformal method will be used later on,
all the physical fields will be decorated with a tilde to
distinguish them from conformally rescaled ones.

\smallskip

The system \eqref{gbu} is a simple extension of a model (the good-bad
subsystem) discussed in \cite{LinRod03} which an example of a system
of wave equations which fails to satisfy the classical null condition
but satisfies the weak null condition. The importance of the good-bad
subsystem is that the non-linearities in this model mimic those
present in the Einstein field equations in harmonic gauge. On the
other hand, the ugly equation models the type of modification of the
standard harmonic gauge formulation of general relativity (by addition
of multiples of the constraint equations) which is needed for damping
violations of the constraints in free numerical evolution schemes
---see \cite{GasHil18}.  Although in \cite{DuaFenGasHil21} the GBU
model was studied in asymptotically flat spacetime backgrounds, for
the purposes of a clean-cut comparison with analogous results obtained
with conformal methods, only the simplest scenario will be considered:
the GBU-system in flat spacetime.  The relevant result of
\cite{DuaFenGasHil21} (with the notation adapted to that of the
present article) for the upcoming comparison is the following:

\begin{proposition}\label{thmASH_flat}
  Let~$(\mathbb{R}^4, \tilde{\bmeta})$ denote the Minkowski spacetime
  and consider a Cartesian coordinate system
  $X^{\alpha}=(\tilde{t},\tilde{X}^{i})$ and let $\{\tilde{u},
  \tilde{\rho},\vartheta^A\}$ denote a coordinate system where
  $\tilde{u}$ is the retarded time, $\tilde{\rho}$ a radial coordinate
  and $\vartheta^A$ some coordinates on $\mathbb{S}^2$. Then, outside
  a compact ball centred at $\tilde{\rho}=0$, the
  \textit{good-bad-ugly} system defined as in equation \eqref{gbu},
  admits formal polyhomogeneous asymptotic solutions near null
  infinity of the form:
  \begin{subequations}
  \begin{align}
    &\tilde{\phi}_g= \sum_{n=1}^{\infty}\frac{G_n}{\tilde{\rho}^n}\label{goodashflat},\\
    &\tilde{\phi}_b= \sum_{n=1}^\infty \frac{B_n}{\tilde{\rho}^n} \label{badashflat},\\
    &\tilde{\phi}_u= \frac{m_{u,1}}{\tilde{\rho}} + \sum_{n=2}^\infty \frac{U_n}{\tilde{\rho}^n}\label{uglyashflat},
  \end{align}
  \end{subequations}
  with ~$G_{n}=G_{n,0}$, $B_n=B_{n,0}+ B_{n,1} \log \tilde{\rho}$
  and~$U_n=U_{n,0}+ U_{n,1}\log \tilde{\rho}$ where each coefficient
  $\Phi_{n,s}$ is constant along outgoing null curves ---namely, they
  are of the form $F(\tilde{u}_{\star},\vartheta^A_{\star})$--- and
  with initial data on~${\mathcal{S}}$ of the form
  \begin{align}\label{eq:initialdata}
    \begin{cases}
      \tilde{\phi}_g\rvert_{\mathcal{S}}=
      \sum_{n=1}^{\infty}\frac{m_{g,n}}{\tilde{\rho}^n}\\ \tilde{\phi}_b\rvert_{\mathcal{S}}=
      \sum_{n=1}^\infty
      \frac{m_{b,n}}{\tilde{\rho}^n}\\ \tilde{\phi}_u\rvert_{\mathcal{S}}=
      \sum_{n=1}^\infty \frac{m_{u,n}}{\tilde{\rho}^n}
    \end{cases}\,,
    \begin{cases}
      \nabla_{\tilde{t}}g\rvert_{\mathcal{S}}=
      \mathcal{O}(\tilde{\rho}^{-2})\\ \nabla_{\tilde{t}}b\rvert_{\mathcal{S}}=
      \mathcal{O}(\tilde{\rho}^{-2})\\ \nabla_{\tilde{t}}u\rvert_{\mathcal{S}}=
      \mathcal{O}(\tilde{\rho}^{-2})
    \end{cases}\,,
	\end{align}
  where~$m_{\phi,n}=m_{\phi,n}(\vartheta^A)$ and $\mathcal{S}$ is the
  hypersurface $\tilde{t}=0$.
\end{proposition}

Observe that according to Proposition \ref{thmASH_flat} the asymptotic
expansion for the good field \eqref{goodashflat} does not contain
logarithmic terms. As it will be explained later, this is not true
since the development of logarithmic terms in the
good field is not excluded by the class of initial data encoded in
equation \eqref{eq:initialdata} ---see Corollaries
\ref{coro:logfree_data_physgood} and \ref{coro:example_pure_log_data}.

\section{The $i^0$-cylinder representation of the Minkowski spacetime}\label{sec:i0cylinder}

Let~$(\tilde{\mathcal{M}},\tilde{\bmeta})$ denote the physical
Minkowski spacetime and consider a spherical polar coordinate system
$(\tilde{t},\tilde{\rho},\vartheta^A)$ where $\vartheta^A$ with
$A=1,2$ are coordinates on $\mathbb{S}^2$ so that one has
\begin{align}\label{eq:MinkowskiMetricPhysicalPolar}
\tilde{\bmeta}=-\mathbf{d}\tilde{t}\otimes\mathbf{d}\tilde{t}
+\mathbf{d}\tilde{\rho}\otimes \mathbf{d}\tilde{\rho}+\tilde{\rho}^2
\mathbf{\bm\sigma},
\end{align}
where~$\bm\sigma$ denotes the standard metric on~$\mathbb{S}^2$ and
~$\tilde{t}\in(-\infty, \infty)$ and $\tilde{\rho}\in [0,\infty)$.
  Let $\{L,\uL,\tilde{\bme}_{\bmA}\}$ denote the following $\tilde
  \bmeta$-null (physical) frame:
\begin{align}
L = \bm\partial_{\tilde{t}} +\bm\partial_{\tilde{\rho}}, \qquad \uL =
\bm\partial_{\tilde{t}}-\bm\partial_{\tilde{\rho}}, \qquad
\tilde{\bme}_{\bmA}=\tilde{\rho}^{-1}\bme_{\bmA},
\end{align}
where $\bme_{\bmA}$ denotes a complex null frame on $\mathbb{S}^2$
with dual coframe $\bm\omega^{\bmA}$ normalised so that
\begin{align}
\bm\sigma=2(\bm\omega^{1}\otimes
\bm\omega^{2}+\bm\omega^{2}\otimes \bm\omega^{1}).
\end{align}
Observe that physical frame is normalised according to
\begin{align}
\tilde{\eta}_{ab}L^a\uL^b
=-\tilde{\eta}_{ab}\tilde{e}^{a}_{1}\tilde{e}^{b}_{2}
=-2.
\end{align} 
Introduce the $F$-coordinate system $(\tau,\rho,\vartheta^A)$ 
  related to the physical coordinates via
\begin{align}\label{eq:Ftophys}
\tau = \frac{\tilde{t}}{\tilde{\rho}}, \qquad \rho =
\frac{\tilde{\rho}}{\tilde{\rho}^2-\tilde{t}^2}.
\end{align}
Additionally, notice that the physical retarded and advanced times
defined as $\tilde{u}:=\tilde{t}- \tilde{\rho}$ and $ \tilde{v}:= \tilde{t}+
\tilde{\rho}$ respectively,
are related to the $F$-coordinates via
\begin{align}\label{eq:UnphysPhysAdvRet}
\tilde{v}=-\rho^{-1}(1-\tau)^{-1}, \qquad \tilde{u}=-\rho^{-1}(1+\tau)^{-1}.
\end{align}
Using equations \eqref{eq:Ftophys} and \eqref{eq:MinkowskiMetricPhysicalPolar}
one readily identifies the following conformal representation of the
Minkowski spacetime $(\mathcal{M},\bmg)$ with
\begin{align}
  \bmg=\Theta^2\tilde{\bm\eta},
\end{align}
where 
\begin{align}\label{eq:i0cylmetricandconffactor}
   \Theta =\rho (1-\tau^2), \qquad
\bmg=- \mathbf{d}\tau\otimes \mathbf{d}\tau
+\frac{(1-\tau^2)}{\rho^2}\mathbf{d}\rho \otimes \mathbf{d}\rho -
\frac{\tau}{\rho} \mathbf{d}\rho\otimes \mathbf{d}\tau -
\frac{\tau}{\rho} \mathbf{d}\tau \otimes \mathbf{d}\rho + \bmsigma.
\end{align}
The unphysical spacetime $(\mathcal{M},\bmg)$  is called
the $i^0$-cylinder representation of the Minkowski spacetime due to
the fact that spatial infinity $i^0$ is ``blown-up'' to an extended set
denoted as $I$ and determined by
\begin{align*}
 I \equiv \{ p \in \mathcal{M} \; \rvert \;\; |\tau(p)|<1, \;
 \rho(p)=0\}. 
\end{align*}
In more general settings there is a residual gauge freedom in the
construction of the $i^0$-cylinder which controls the way in which the
cylinder intersects $\mathscr{I}^{\pm}$.  The current construction
corresponds to the the \emph{horizontal} representation of the
$i^0$-cylinder as future and past null infinity are located at:
\begin{align}
 \mathscr{I}^{+} \equiv \{ p \in \mathcal{M} \; \rvert\; \tau(p) =1
 \}, \qquad \mathscr{I}^{-} \equiv \{ p \in \mathcal{M} \; \rvert \;
 \tau(p) =-1\}.
\end{align}
and the \emph{critical sets} where $\mathscr{I}^{\pm}$ meet the $i^0$-cylinder correspond to
\begin{align*}
 I^{+} \equiv \{ p\in \mathcal{M} \; \rvert \; \tau(p)=1, \; \rho(p)=0
 \}, \qquad I^{-} \equiv \{p \in \mathcal{M}\; \rvert \; \tau(p)=-1,
 \; \rho(p)=0\}.
\end{align*}
To round up the discussion, one introduces a $\bmg$-null frame 
$\{\bme, \bmue, \bme_{\bmA}\}$ where
\begin{align}\label{eq:Fframe}
 \bme =(1+\tau)\bm\partial_{\tau}{} - \rho\bm\partial_{\rho}{}, \qquad
 \bmue =(1-\tau)\bm\partial_{\tau}{} + \rho\bm\partial_{\rho}{},
\end{align}
which will be called the $F$-frame.  Notice that the unphysical frame is
$g$-normalised so that
\begin{align*}
  g_{ab}e^a\ue^b=-g_{ab}e_{1}^ae_{2}^{b}=-2.
\end{align*}
In the up-coming discussion yet another set of null-frames will be relevant, which are the NP
frame adapted to $\mathscr{I}^{\pm}$, respectively. The relation between these frames
is given in the following:

 \begin{proposition}\label{Prop:NPtoFgauge}
   The $\mathscr{I}^{\pm}$-NP-frame, the $F$-frame and the
   standard physical frame for the Minkowski spacetime are related via:
\begin{align*}
  \text{\emph{NP hinged at} $\mathscr{I}^{+}$}:& \quad\bme^{+} =
  4(\Lambda_{+})^{2} \bme = \Theta^{-2} L, && \bmue^{+}=
  \tfrac{1}{4}(\Lambda_{+})^{-2}\bmue = \uL, && \bme_{\bmA}^{+}=
  \bme_{\bmA}= \Theta^{-1}\tilde{\bme}_{\bmA}\\ \text{\emph{NP hinged
    at} $\mathscr{I}^{-}$}:& \quad\bme^{-} =
  \tfrac{1}{4}(\Lambda_{-})^{-2} \bme = L, && \bmue^{-}=
  4(\Lambda_{-})^{2}\bmue = \Theta^{-2} \uL, && \bme_{\bmA}^{-}=
  \bme_{\bmA}= \Theta^{-1}\tilde{\bme}_{\bmA}.
\end{align*}
where  the conformal factor $\Theta$ and
Lorentz transformation (encoded via the boost parameter $\varkappa$)
relating these frames are given by  
\begin{align}\label{eq:CF-thetaAndBoostParameter}
  &\Theta := \rho (1-\tau^2) = \frac{1}{\tilde{\rho}},  
   && \varkappa  := \frac{1+\tau}{1-\tau} = -\frac{\tilde{v}}{\tilde{u}}.
  \\
  & (\Lambda_{+})^{2}:= \Theta^{-1}\varkappa^{-1}=
  \rho^{-1}(1+\tau)^{-2},  && (\Lambda_{-})^{2}:= \Theta^{-1}\varkappa=
  \rho^{-1}(1-\tau)^{-2}.
\end{align}
 \end{proposition}
 For further discussion on the $i^0$-cylinder
 see \cite{DuaFenGasHil22b, MinMacVal22, Val16, Fri02} and references there in,
 and for the relation between  the frames see
 \cite{GasVal20, GasPin23, FriKan00}.

 \section{The unphysical GBU model}\label{sec:GBUunphysical}

 To derive the unphysical (conformal) counterpart
 of the GBU model, one starts by recalling that, given
 any two conformally related metrics $\bmg = \Omega^2 \tilde{\bmg}$, the
 D'Alembertian operator (acting on scalars) transforms as
\begin{align}
  \square \phi- \frac{1}{6} \phi R = \Omega ^{-3} \bigg(
  \tilde{\square }\tilde{\phi}- \frac{1}{6} \tilde{\phi}
  \tilde{R}\bigg), \label{eq:General_Wave_Conformal_Transformation}
\end{align}
where $\phi=\Omega^{-1}\tilde{\phi}$.
The key observation is that for the
$i^0$-cylinder Minkowski spacetime
---see equation \eqref{eq:i0cylmetricandconffactor}---
one has $R=0$. Moreover, since $\tilde{R}=0$ then equation
\eqref{eq:General_Wave_Conformal_Transformation}
reduces to
\begin{align} 
    \square \phi = \Theta ^{-3}
  \tilde{\square }\tilde{\phi}.
\end{align}
Hence, the unphysical good equation is given simply by
\begin{align}\label{eq:Unphysical_wave}
  \square \phi =0.
\end{align}
Observe that for other conformal transformation for which $R \neq 0$
the resulting unphysical good equation would be formally singular at this level
due to the presence of negative powers of the conformal factor.
An additional remark to be made is that  that the associated unphysical field
$\phi=\Theta^{-1}\tilde{\phi}$ corresponds to the radiation field since, expressed
 in physical coordinates, $\Theta = \tilde{\rho}^{-1}$. This
is not true for other conformal representation of
Minkowski spacetime.

\medskip

The last discussion shows how to obtain the unphysical good equation.
A similar calculation exploiting the relations given in
Section \ref{sec:i0cylinder}  shows that the unphysical GBU-system
is given by
\begin{subequations}\label{unphysical_GBU}
\begin{align}
   \square \phi_g & =0, \label{unphysical_good} \\
   \square \phi_b &= \frac{1}{4}\Theta ( \varkappa \nabla _{\bmue}\phi_g +
   \varkappa^{-1}\nabla _{\bme}\phi_g)^2,  \label{unphysical_bad}\\   
  \square \phi_u & =   \varkappa \nabla _{\bmue}\phi_u +
  \varkappa^{-1}\nabla _{\bme}\phi_u,  \label{unphysical_ugly}
\end{align}
\end{subequations}
Notice that, unlike the unphysical good equation,
the unphysical bad and ugly equation become formally singular
due to the presence of the boost parameter $\varkappa$ and its inverse
which diverge at $\mathscr{I}^{\pm}$, respectively.

\medskip

A common feature of several linear model equations on the
$i^0$-cylinder Minkowski background is that, in general, the principal
symbol of the equation looses rank at the critical sets. In the
present case this can be observed from writing the unphysical good
equation \eqref{unphysical_good} explicitly in $F$-coordinates:
\begin{align}\label{eq:UnphysicalWaveExplicit}
  (\tau ^2-1) \partial _{\tau}^2 \phi_g -2 \rho \tau \partial
  _{\tau}\partial _{\rho}\phi_g + \rho ^2 \partial _{\rho}^2\phi_g + 2
  \tau \partial _{\tau}\phi_g + \Delta _{\mathbb{S}^{2}{}}{}\phi_g =
  0,
\end{align}
where~$\Delta _{\mathbb{S}^{2}{}}{}$ is the Laplace operator on
$\mathbb{S}^2$. In other words, if one divides the the equation by
$(\tau ^2-1)$ to isolate the highest $\tau$-derivatives one obtains a
formally singular equation.  This situation is not ameliorated for the
unphysical ugly and bad equations where the formally singular terms
appear even in lower order terms (not entering the principal part) of
the equation ---see equation \eqref{unphysical_GBU}.  Nevertheless,
one can still formally solve these equations as shown below.
Furthermore, in some cases (such as the spin-2 field) the formal
solutions built with this method have been satisfactorily established
by proving estimates that remain valid up to the critical sets as in
\cite{Fri02} ---see also \cite{OliOlv21, TauVal23}.  The focus of this
article is on the construction of formal solutions with conformal
methods for deriving higher order expansions similar to that of
\cite{DuaFenGasHil21, DuaFenGas22, DuaFenGasHil22a} but which capture
missed features (logarithmic terms) by the analysis in the latter
references.  Obtaining estimates for the good-bad-ugly model will be
left for future work.

\medskip

Although the ugly and bad field contain further singular terms, the
key equation to be solved is the good equation. To do so, consider the
Ansatz :
\begin{align}\label{eq:Ansatz}
\phi= \sum_{p=0}^{\infty}\sum_{\ell=0}^{p}\sum_{m=-\ell}^{m=\ell}
\frac{1}{p!}a_{p;\ell,m}(\tau)Y_{\ell m}\rho^{p},
\end{align}
where $Y_{\ell,m}$ are the spherical harmonics.
Taking this Ansatz already restricts the initial data to be analytic
close to $i^0$ as:
\begin{align}\label{eq:ID_field}
\phi|_{\mathcal{S}} = \sum_{p=0}^{\infty}
\sum_{\ell=0}^{p}\sum_{m=-\ell}^{m=\ell}\frac{1}{p!}a_{p;\ell,m}(0)Y_{\ell
  m}\rho^p, \qquad \dot{\phi}|_{\mathcal{S}} =
\sum_{p=0}^{\infty}\sum_{\ell=0}^{p}\sum_{m=-\ell}^{m=\ell}
\frac{1}{p!}\dot{a}_{p;\ell,m}(0)Y_{\ell m}\rho^p,
\end{align}
where derivatives respect to $\tau$ are denoted as an over-dot.
Notice from equation \eqref{eq:Ftophys} that for $\tilde{\rho}\neq 0$,
the initial hypersurface $\mathcal{S}$ can be thought as the
$\tau=0$ hypersurface in the unphysical picture or the $\tilde{t}=0$
in the physical one. The crucial advantage of using the
Ansatz \eqref{eq:Ansatz} is that that solving
\eqref{eq:UnphysicalWaveExplicit} reduces to solving the ordinary
differential equation (ODE)
\begin{align}\label{ODE_wave_JacobiPoly}
(1-\tau^2)\ddot{a}_{p;\ell,m} +
  2\tau(p-1)\dot{a}_{p,\ell,m}+(\ell+p)(\ell-p+1){a}_{p;\ell,m}=0,
\end{align}
for $a_{p;\ell,m}(\tau)$. Equation \eqref{ODE_wave_JacobiPoly} is a
Jacobi equation whose solutions are well-known ---see \cite{Sze78}.
Hence, the solution to the unphysical good field is given by
substituting into the Ansatz \eqref{eq:Ansatz}, for each $p,\ell,m$
the following expressions for $a_{p;\ell,m}(\tau)$:

\begin{lemma}[wave equation on the $i^0$-cylinder
    background~\cite{MinMacVal22}]\label{Lemma:Sol_Jacobi_and_Logs}
  The solution to the Jacobi differential equation implied by
  the Ansatz \eqref{eq:Ansatz} and equation
  \eqref{unphysical_good} is given
   by:
  \begin{enumerate}[label=(\roman*)]
  \item For $p\geq 1$   and $0\leq \ell \leq p-1$
    \begin{align}\label{eq:Sol_jac_poly}
      a(\tau)_{p;\ell,m} =A_{p,\ell,m}
      \bigg(\frac{1-\tau}{2}\bigg)^{p}
      J_{\ell}^{(p,-p)}(\tau) + B_{p,\ell,m}
      \bigg(\frac{1+\tau}{2}\bigg)^{p}J_{\ell}^{(-p,p)}(\tau)
    \end{align}
  \item For  $p\geq 0$   and $\ell=p$:
    \begin{align}\label{eq:Sol_highestharmonic}
      {a}_{p;p,m}(\tau) = \bigg(\frac{1-\tau}{2}\bigg)^{p}
      \bigg(\frac{1+\tau}{2}\bigg)^{p}\Bigg(C_{p,p,m} +D_{p,p,m}
      \int_{0}^{\tau} \frac{ds}{(1-s^2)^{p+1}}\Bigg)
    \end{align}
    where ~$A_{p,\ell,m}$, $B_{p,\ell,m}$, $C_{p,p,m} $ and $D_{p,p,m} $
    are constants determined by the
    initial data as given in \eqref{eq:ID_field} implicitly through
    ~$a_{p;\ell,m}(0)$ and $\dot{a}_{p;\ell,m}(0)$ and
    ~$J^{\alpha,\beta}_{\gamma}(\tau)$ are the Jacobi polynomials.
  \end{enumerate}
\end{lemma}
\noindent The solutions for $p=\ell$ are not only polynomials but contain
logarithmic terms as direct inspection for a few values of $p$ reveals:
\begin{align}
  {a}_{0;0,0}(\tau) & = C_{000} + \tfrac{1}{2} D_{000} \log \varkappa,
  \\ {a}_{1;1,m}(\tau) & = \tfrac{1}{4} (1 - \tau )
  (1 + \tau ) (C_{11m} + \tfrac{1}{4} D_{11m} ( \log \varkappa  + 2\tau(1-\tau^2))).
\end{align}
Interestingly, these logarithmic terms enter linearly at each order
and appear in combinations that can be expressed in terms of the boost
parameter $\varkappa:=(1+\tau)/(1-\tau)$, which in physical
coordinates corresponds to the quotient $-\tilde{v}/\tilde{u}$.  For
linear fields such as the spin-1 and spin-2 fields propagating in the
$i^0$-cylinder Minkowski background, these logarithmic terms are
responsible for the polyhomogeneous peeling behaviour of the field
---see \cite{MinMacVal22, ValAli22}.  For the non-linear case of the
(conformal) Einstein field equations further obstructions to the
smoothness of $\mathscr{I}$ arise ---see \cite{Val04}.  To see how the
regularity of the field is controlled by the initial data observe that
\begin{align}\label{DtoID}
D_{p,p,m}=2^p \dot{a}_{p,p,m}(0),
\end{align}
hence, choosing initial data such that $\dot{a}_{p,p,m}(0)=0$ one
obtains a solution without the logarithmic mentioned above. This is
summarised in the following:
\begin{lemma}\label{Lemma:logfreeRemark}(Regularity
  condition~\cite{MinMacVal22}).
   If initial data is chosen so that $\dot{a}_{p;p,m}(0)=0$ then the
   solution extends analytically to the critical sets.
\end{lemma}

To translate these results to the physical picture
observe that using $\tilde{\phi}_g=\Theta \phi_g$ and equations
\eqref{eq:Ansatz}, \eqref{eq:UnphysPhysAdvRet} and Lemma
\ref{Lemma:Sol_Jacobi_and_Logs} one can obtain an expression for the
physical good field. Furthermore, to be in the same set-up of the
asymptotic expansions of \cite{DuaFenGasHil21, DuaFenGas22,
  DuaFenGasHil22a}, one needs to evaluate the field along an outgoing
null curve $\gamma =
(\tilde{u}_{\star},\tilde{\rho},\vartheta^A_{\star})$ where
$\tilde{u}_{\star}$ and $\vartheta^A_{\star}$ are constant.
In doing so, one obtains:
\begin{align}\label{eq:goodexpansionwithlogs}
     \tilde{\phi}_{g} \simeq
     \sum_{\ell=0}^{\infty}\sum_{m=-\ell}^{\ell}\alpha_{\ell}(\tilde{u_\star},
     \vartheta^A_{\star}) D_{\ell \ell m}\frac{\log
       \tilde{\rho}}{\tilde{\rho}^{\ell+1}} +
     \sum_{n=1}^{\infty}\frac{G_{n}(\tilde{u}_\star,
       \vartheta^A_{\star})}{\tilde{\rho}^{n}}
\end{align}
where $\alpha_{\ell}$ and $G_{n}$ are constant along $\gamma$.
To understand how the regularity condition of Lemma
\eqref{Lemma:logfreeRemark} is expressed in the physical picture observe
that splitting the initial data for~$\dot{\phi}$ as
\begin{align}
  \dot{\phi}|_{\mathcal{S}} =\sum_{p=0}^{\infty}
  \sum_{m=-p}^{m=p}\frac{1}{p!}\dot{a}_{p;p,m}(0)Y_{\ell m}\rho^p +
  \sum_{p=1}^{\infty}\;\;\sum_{\ell=1}^{\max\{1, \; p-1\}}
  \sum_{m=-\ell}^{m=\ell}\frac{1}{p!}\dot{a}_{p;\ell,m}(0)Y_{\ell
    m}\rho^p ,
\end{align}
and translating to the physical set-up using that
$\dot{\phi}|_{\mathcal{S}} =
\tilde{\rho}^2\partial_{\tilde{t}}\tilde{\phi}|_{\mathcal{S}}$ one
obtains the following
\begin{corollary}\label{coro:logfree_data_physgood}
  \emph{Initial data for the physical good field satisfying
\begin{align}\label{eq:finetunedID}
\tilde{\phi}|_{\mathcal{S}} =
\sum_{p=0}^{\infty}\sum_{\ell=0}^{p}\sum_{m=-\ell}^{m=\ell}
\frac{1}{p!}a_{p;\ell,m}(0)Y_{\ell m}\tilde{\rho}^{-p-1}, \quad
\partial_{\tilde{t}}\tilde{\phi}|_{\mathcal{S}}=
\sum_{p=1}^{\infty}\sum_{\ell=1}^{{\max\{1, \; p-1\}}}
\sum_{m=-\ell}^{m=\ell}\frac{1}{p!}\dot{a}_{p;\ell,m}(0)Y_{\ell m}
\tilde{\rho}^{-p-2},
\end{align}
gives rise to a log-free expansion close to $I$.  }
\end{corollary}
Moreover, to further emphasise the relation between initial data and the
development of logarithmic terms one can construct a solution with
$A_{p,\ell,m}=B_{p,\ell,m}=C_{p,\ell,m}=0$ but $D_{p,\ell,m} \neq 0$
by considering initial data such that
\begin{corollary}\label{coro:example_pure_log_data}
   \emph{Initial data for the physical good field satisfying
\begin{align}
\tilde{\phi}|_{\mathcal{S}} =0, \qquad
\partial_{\tilde{t}}\tilde{\phi}|_{\mathcal{S}}=
\sum_{p=0}^{\infty}\sum_{m=-p}^{m=p}\frac{1}{p!}\dot{a}_{p;p,m}(0)Y_{p
  m} \tilde{\rho}^{-p-2},
\end{align}
gives rise to an solution with
$A_{p,\ell,m}=B_{p,\ell,m}=C_{p,\ell,m}=0$ but $D_{p,\ell,m} \neq 0$
hence containing logarithmic terms. }
\end{corollary}
Although Corollaries \ref{coro:logfree_data_physgood} and
\ref{coro:example_pure_log_data} are simple implications of Lemma
\ref{Lemma:logfreeRemark} translated to the physical set-up, they
clearly show that, contrary to what one would naively expect, the
development of the logarithmic terms is not simply associated to the
rough decay of the initial data close to $i^0$. Namely, generic
initial data (meaning not fine-tuned as that of equation
\eqref{eq:finetunedID}) with
$\partial_{\tilde{t}}\tilde{\phi}|_{\mathcal{S}} \simeq
\mathcal{O}(\tilde{\rho}^{-p-2})$ for some $p \in
\mathbb{N}_0$ will produce log terms starting  at order
$\tilde{\phi} \simeq \tilde{\rho}^{-p-1}\log \tilde{\rho} $.  On the
other hand, initial data with the same rough decay but fine-tuned
---see Corollary \eqref{coro:logfree_data_physgood}--- produces
log-free expansions.

\medskip

The above discussion for the good field constitutes the basic
construction for obtaining the solution for the bad and ugly fields. To
see this observe that, once the solution for the good field is known,
the source term in the bad equation is fixed:
\begin{align}\label{wave_inhomogeneous}
  \square \phi = f,
\end{align}
where~$f=f(\tau,\rho,\theta^A)$ is a given source. 
Then,  when considering the analogous Ansatz \eqref{eq:Ansatz}
for the bad field $\phi_b$, one obtains an ODE
\begin{align}\label{ODE_wave_JacobiPoly_source}
(1-\tau^2)\ddot{a}_{p;\ell,m} +
  2\tau(p-1)\dot{a}_{p,\ell,m}+(\ell+p)(\ell-p+1){a}_{p;\ell,m}
  =f_{p;\ell,m}(\tau),
\end{align}
where~$f_{p;\ell,m}(\tau)$ arises from expanding $f$ according to the
Ansatz~\eqref{eq:Ansatz}. 
Equation \eqref{ODE_wave_JacobiPoly_source}
can be solved by the method of variation of parameters (similar to the standard
Duhamel's principle for the wave equation in Minkowski spacetime) and
the following can be derived:
\begin{lemma}[Inhomogeneous wave equation on the $i^0$-cylinder
    background~\cite{MinMacVal22}]\label{ODE_bad_sol} The solution to
  the Jacobi differential equation implied by the Ansatz
  \eqref{eq:Ansatz} and equation \eqref{wave_inhomogeneous} is given by:
\begin{align}
  a_{p;\ell,m}(\tau)=a^{H}_{1:p;\ell,m}(\tau)b_{1:p;\ell,m}(\tau) +
  a^{H}_{2:p;\ell,m}(\tau)b_{2:p;\ell,m}(\tau),
\end{align}
where~$a^{H}_{1:p;\ell,m}(\tau)$ and $a^{H}_{2:p,\ell,m}(\tau)$ are
two independent solutions to the homogeneous problem as given in Lemma
\ref{Lemma:Sol_Jacobi_and_Logs} while~$b_{1:p;\ell,m}(\tau)$ and
$b_{2:p;\ell,m}(\tau)$ are given by
\begin{subequations}\label{eq:varpar}
\begin{align}
 & b_{1:p;\ell,m}(\tau)= F_{1;p;\ell,m} -
  \int_{0}^{\tau}\frac{a_{2:p;\ell,m}(s)f_{p,\ell,m}(s)}{W_\star(1-s^2)^{p}}ds,
  \\ & b_{2:p,\ell,m}(\tau)= F_{2; p;\ell,m} -
  \int_{0}^{\tau}\frac{a_{1:p;\ell,m}(s)f_{p,\ell,m}(s)}{W_\star(1-s^2)^{p}}ds.
\end{align}
\end{subequations}
where~$F_{1,p;\ell,m}$, $F_{2,p;\ell,m}$ and~$W_\star$ are constants.
\end{lemma}
From equation \eqref{eq:varpar} one can examine how behaviour of the
source affects the solution. In particular if the source
$f_{p;\ell,m}(\tau)$ contains logarithmic terms then
$b_{1:p;\ell,m}(\tau)$ can develop terms of the form
$(\log \varkappa )^n$ for some $n \in \mathbb{N}$,
as it is the case 
for the Maxwell-scalar field system in \cite{MinMacVal22}.
Surprisingly, in the case of equation \eqref{unphysical_bad}, despite
that $\phi_{g}$ contains logarithmic terms, the associated source to
equation \eqref{unphysical_bad} does not. The reason for this is can
be more clearly seen in the physical picture from the fact that the
source term is $(\partial_{\tilde{t}}\tilde{\phi}_{g})^2$ and that
$\partial_{\tilde{t}} \log \varkappa
=(\tilde{v}^{-1}-\tilde{u}^{-1})/2$.  Proceeding as before, writing
the corresponding physical solution along outgoing null curves
$\gamma$ one obtains
\begin{align}\label{bad_expansion}
    \tilde{\phi}_b \simeq
    \sum_{\ell=0}^{\infty}\beta_{\ell}(\tilde{u}_{\star},\vartheta^A_{\star})
    \frac{\log \tilde{\rho}}{\tilde{\rho}^{\ell+1}} +
    \sum_{n=1}^{\infty}\frac{\mathcal{B}_{n}(\tilde{u}_{\star},\vartheta^A_{\star})}{\tilde{\rho}^{n}}.
  \end{align}
Observe that
the logarithmic terms again enter only linearly. The coefficient
$\beta_{\ell}$ contains terms coming not only from the logarithmic
part of the good field (that controlled by the coefficient
$D_{\ell;\ell,m}$) but also coefficients coming from other parts of
the initial data for the good such as $A_{p;\ell,m}$, $B_{p;\ell,m}$
and $C_{p;\ell,m}$. Additionally, it depends on the initial data for
the bad field encoded through the coefficients $F_{1;p;\ell,m}$ and
$F_{2;p,\ell,m}$ appearing in equation
\eqref{eq:varpar}. Although in principle
it could be possible to investigate the condition controlling
appearance the logarithmic terms in equation \eqref{bad_expansion} in terms of
the initial data for the bad field and the good field (comprising the
source) one would obtain convoluted expressions (at each order in
$p$) an such analysis was out of the scope of \cite{DuaFenGasHil22b} and this
review.  This situation in stark contrast with the solution for the
good where the appearance of the logarithmic terms can be cleanly
controlled at the level of initial data through the coefficient
$D_{p;p,m}$.

\medskip

For the ugly field the situation is different since
equation \eqref{unphysical_ugly} is
not of the form of expression \eqref{wave_inhomogeneous}. Nevertheless,
one can still exploit the solution to the good field to obtain a
solution to the ugly field. To do so, it is better to consider the
equations in the physical set-up. Exploiting the following identity
\begin{align}\label{commutation_rewrriten}
\underline{L}\Bigg(\tilde{\rho}^2\Big( \tilde{\square} \tilde{\phi} -
\frac{2}{\tilde{\rho}}\nabla_{\tilde{t}}\tilde{\phi})\Big)
\Bigg)
=\tilde{\rho}\tilde{\square}(\tilde{\rho}\underline{L}\tilde{\phi}),
\end{align}
and introducing an auxiliary field
$\tilde{\Phi}:=\tilde{\rho}\uL \tilde{\phi}_u$, it can be shown that 
$\tilde{\square}\tilde{\Phi}=0$ implies a solution to
\begin{align}
\tilde{\square} \tilde{\phi} -
\frac{2}{\tilde{\rho}}\nabla_{\tilde{t}}\tilde{\phi}=Q,
\end{align}
with $\uL(\tilde{\rho}^2Q)=0$. The initial data for $\tilde{\Phi}$
and $\tilde{\phi}$ are related via:
\begin{subequations}
  \begin{align}
    \tilde{\Phi}|_{\mathcal{S}} & =
    \left[\tilde{\rho}(\partial_{\tilde{t}}-\partial_{\tilde{\rho}})
    \tilde{\phi}\right]_{\mathcal{S}},\\ 
    \partial_{\tilde{t}}\tilde{\Phi}|_{\mathcal{S}} &=
    \left[\tilde{\rho}(\partial^{2}_{\tilde{t}}
    -\partial_{\tilde{\rho}}\partial_{\tilde{t}})\tilde{\phi}\right]_{\mathcal{S}}
    = \left[\tilde{\rho}(\Delta \tilde{\phi} + 2
    \tilde{\rho}^{-1}\partial_{\tilde{t}}\tilde{\phi}-Q
    -\partial_{\tilde{\rho}}\partial_{\tilde{t}}\tilde{\phi})\right]_{\mathcal{S}}
  \end{align}
\end{subequations}
Then, providing appropriately the initial data ---so that $Q=0$--- one can
then construct a solution to equation \eqref{unphysical_ugly}
from the solution to
\eqref{unphysical_good}. Exploiting the solution for the good field $\phi_g$
in the unphysical picture as encoded in Lemma
\ref{Lemma:Sol_Jacobi_and_Logs},
a calculation gives that along $\gamma$ :
 \begin{align*}
    \tilde{\phi}_u \simeq \sum_{\ell=0}^{\infty}w_{\ell}(\tilde{u}_\star,\vartheta^A_{\star})
     \frac{\log \tilde{\rho}}{\tilde{\rho}^{\ell+1}} +
\sum_{n=1}^{\infty}\frac{\mathcal{W}_{n}(\tilde{u}_\star,\vartheta^A_{\star})}{\tilde{\rho}^{n}}.
 \end{align*}
 As in the case of the bad field, establishing how the coefficients
 $w_{\ell}(\tilde{u}_\star,\vartheta^A_{\star})$ are determined in
 terms of initial data will not be pursued here.

 \medskip

 The analysis made for the GBU model close to spatial and null
 infinity in Minkowski spacetime is enough to to show that the
 logarithmic terms discussed in~\cite{DuaFenGasHil21} cannot
 correspond to those logarithmic terms having origin at the critical
 sets using the~$i^0$-cylinder framework. This is shown categorically
 through the analysis of the good field: the expansion for the good
 field as reported in\cite{DuaFenGasHil21} is log-free while the one
 obtained using the $i^0$-cylinder framework it is not unless the
 initial data is fine-tuned. In the following section, the asymptotic system
 method of \cite{DuaFenGasHil21} is revisited to see if a
 modified version of the asymptotic system heuristics can incorporate
 the logarithmic terms in the good field and if such terms can be put
 in correspondence with those of the unphysical set-up. Interestingly,
 the construction of such asymptotic expansions which are sensitive to the
 $i^0$-cylinder logs are related to a set of conservation laws related
 to the so-called the Newman-Penrose constants.

\section{Conservation laws and asymptotic system expansions}\label{sec:conslaws}

The asymptotic expansions of \cite{DuaFenGasHil21, DuaFenGas22,
  DuaFenGasHil22a} are based on the notion of H\"ormander's asymptotic
system and the weak null condition of \cite{LinRod03},
hence, the heuristic guiding principle of
 of \cite{DuaFenGasHil21, DuaFenGas22, DuaFenGasHil22a} is that
derivatives tangent to the outgoing null cone  decay faster than
transverse ones. In the asymptotic system heuristics,
the former derivatives are called good derivatives
and the latter are called bad derivatives.
The analysis of Section
\ref{sec:GBUunphysical} shows that the analysis in \cite{DuaFenGasHil21}
missed the $i^0$-cylinder logs captured by conformal methods as
clearly evidenced in the case of the good field by comparing equations
\eqref{goodashflat} and \eqref{eq:goodexpansionwithlogs}.
Therefore, the natural question is if
asymptotic expansions capturing the $i^0$-cylinder logs
can be derived from on a suitably modified
asymptotic system notion.
To do this, recall that the (first order)
asymptotic system is obtained by disregarding the terms
that only contain good derivatives.
Namely, expressing equation $\tilde{\square}\tilde{\phi}=0$
using the physical null frame and discarding terms that
only contain good derivatives one gets
\begin{align}\label{eq:asymptsystemleadingrevisited}
  \uL L (\tilde{\rho}\tilde{\phi}) \simeq 0.
\end{align}
The latter asymptotic equation can be integrated as follows
\begin{align}\label{asymptsysSol}
  \partial_{\tilde{u}}\partial_{\tilde{v}}(\tilde{\rho}\tilde{\phi})
  \simeq & \;0,
  \nonumber\\ \partial_{\tilde{v}}(\tilde{\rho}\tilde{\phi}) \simeq &
  (\partial_{\tilde{v}}(\tilde{\rho}\tilde{\phi}))|_{\tilde{u}_\star}
  :=f(\tilde{v},\vartheta^A),
  \nonumber \\ \tilde{\rho}\tilde{\phi} \simeq &
  (\tilde{\rho}\tilde{\phi})|_{\tilde{v}_\star} +
  \int_{\tilde{v}_\star}^{\tilde{v}}f(\bar{v},\vartheta^A)d\bar{v},
  \nonumber \\ \implies \tilde{\phi} \simeq &
  \frac{1}{\tilde{\rho}}(G(\tilde{u},\vartheta^A)+F(\tilde{v},\vartheta^A)).
\end{align}
Observe that if initial data is provided such that
\begin{align}\label{eq:leadingnolog}
  f(\tilde{v},\vartheta^A)\simeq \tilde{v}^{-1}M(\vartheta^A) + O(\tilde{v}^{-2}),
\end{align}
where $M$ depends only on~$\vartheta^A$
then the integration leading to equation ~\eqref{asymptsysSol} gives
\begin{align}
  \tilde{\phi} \simeq \frac{\log \tilde{v}}{\tilde{\rho}}M(\vartheta^A)
  + \frac{1}{\tilde{\rho}}G(\tilde{u},\vartheta^A),
\end{align}
Therefore, evaluating the field on outgoing null curves
$\gamma$ gives
\begin{align}\label{eq:phifirstlog}
\tilde{\phi} \simeq \frac{\log
  \tilde{\rho}}{\tilde{\rho}}M( \vartheta^A) + \frac{1}{\tilde{\rho}}
G(\tilde{u},\vartheta^A).
\end{align}
Thus, to leading order one can recover a logarithmic term in the
expansion for the good field. However, this still does not show that
this logarithmic term is indeed the same one obtained for $p=0$ by
means of the $i^0$-cylinder construction. To clarify their relation,
recall that $f(\tilde{v},\vartheta):=
(\partial_{\tilde{v}}(\tilde{\rho}\tilde{\phi}))|_{\tilde{u}_\star}$
from which follows that
a necessary and sufficient condition at the level of initial data
to remove the logarithmic term in equation \eqref{eq:phifirstlog}
is:
\begin{align}\label{nologleadID}
\big(\partial_{\tilde{t}}(\tilde{\rho}\tilde{\phi}) +
\partial_{\tilde{\rho}}(\tilde{\rho}\tilde{\phi})\big)|_{\mathcal{S}}
\simeq \mathcal{O}(\tilde{\rho}^{-2}),
\end{align}
Then, expressing equation ~\eqref{nologleadID} in terms the unphysical picture
and substituting the initial data Ansatz~\eqref{eq:Ansatz}  one gets
\begin{align}\label{nologleadIDPhysUnphysRelation}
\big(\partial_{\tilde{t}}(\tilde{\rho}\tilde{\phi}) +
\partial_{\tilde{\rho}}(\tilde{\rho}\tilde{\phi})\big)|_{\mathcal{S}}
\simeq \frac{1}{\tilde{\rho}}D_{000} + \mathcal{O}(\tilde{\rho}^{-2}).
\end{align}
Therefore, the regularity condition at order $p=0$
 controlling the appearance of the leading log
in the unphysical picture corresponds  to the
condition~\eqref{nologleadID} obtained with the (improved)
heuristic asymptotic system method discussed above.

\medskip

Observe that the term $p=0$ in the Ansatz \eqref{eq:Ansatz} only
describes the spherically symmetric part of the solution, and as
emphasised through Corollaries \ref{coro:logfree_data_physgood} and
\ref{coro:example_pure_log_data}, the development of the logarithmic
terms at order $p \geq 1$ are controlled by the coefficients $D_{p p
  m}$ appearing in a spherical harmonic decomposition of the field.
Hence, if the aim is to devise an expansion scheme which makes direct
contact with the ($i^0$-cylinder) logarithmic terms controlled by
$D_{p,p,m}$, one needs to resort to a spherical harmonic decomposition
of the (physical) field. Interestingly, the construction of such
asymptotic expansion scheme is related to a set of conservation
identities giving rise to the Newman-Penrose constants.  The
Newman-Penrose constants \cite{NewPen68} are a set of conserved
quantities at null infinity. For linear fields propagating in
Minkowski spacetime such as the spin-1 (Maxwell's equations)
and spin-2 (linearised gravity) it was shown in \cite{NewPen68}
that there is an infinite hierarchy of these quantities.
However, in the non-linear gravitational case only 10 of these
quantities remain conserved. In general, these conserved quantities
arise as a consequence of asymptotic conservation laws. For the
case of the wave equation (massless spin-0) in flat spacetime
\begin{align}\label{eq:spin0fieldgood}
\tilde{\square}\tilde{\phi}=0,
\end{align}
one has, in fact, the following set of (exact) conservation laws:
\begin{align}\label{eq:cons_laws}
 \uL (\tilde{\rho}^{-2\ell}L (\bme^{+})^{\ell}\psi_{\ell m}) = 0,
 \qquad L(\tilde{\rho}^{-2\ell} \uL (\bmue^{-})^{\ell}\psi_{\ell m}) =
 0,
\end{align}
where $\psi_{\ell m}= \int_{\mathbb{S}^2} \phi \; Y_{\ell m} \;
d\sigma$ with $d\sigma$ denoting the area element in $\mathbb{S}^2$
and $\psi=\tilde{\rho}\tilde{\phi}$ is the radiation field ---see
\cite{Keh21_a, GajKeh22}.
Translating these identities to the
conformal set-up is simple since $\Theta = \tilde{\rho}^{-1}$,
and, hence, the radiation
field coincides with the unphysical field: $\psi =
\phi:=\Theta^{-1}\tilde{\phi}$.
For the $\mathscr{I}^{+}$-adapted set-up, the relevant identity is
the first one in equation \eqref{eq:cons_laws}, which integrating once gives 
\begin{align}
\tilde{\rho}^{-2\ell}L (\bme^{+})^{\ell}\phi_{\ell m} = f_{\ell m} (\tilde{v}).
\end{align}
where each $f_{\ell m}$ is a function that only depends on $\tilde{v}$.
The latter equation can be written more compactly as
\begin{align}\label{eq:high-order-asymptotic-system-good}
\tilde{\rho}^{-2\ell-2} (\bme^{+})^{\ell+1}\phi_{\ell m} = f_{\ell m}
(\tilde{v}).
\end{align}
Furthermore, for $\ell=0$, the exact conservation law of
equation \eqref{eq:cons_laws} reads
\begin{align}\label{eq:asymptlead}
\partial_{\tilde{u}} \partial_{\tilde{v}} \phi_{00}=0,
\end{align}

Observe that expressions \eqref{eq:asymptlead} and
\eqref{eq:asymptsystemleadingrevisited} are similar, however, the
former comes from an exact conservation law while the latter is an
asymptotic relation.  One can consider either of these two expressions
as encoding the \emph{leading order asymptotic system}, nonetheless,
as it will be shown in the remaining of this section expression
\eqref{eq:asymptlead} gives a direct route to derive a \emph{higher
order asymptotic system} which is sensitive to the $i^0$-cylinder logs
controlled by the regularity condition of Lemma
\ref{Lemma:logfreeRemark}.  Integrating equation \eqref{eq:asymptlead}
once gives
\begin{align}
  \partial_{\tilde{v}} \phi_{00}=f_{00}(\tilde{v}),
  \end{align}
with
\begin{align}
  f_{00} (\tilde{v}):=
  \partial_{\tilde{v}}\phi_{00}|_{\tilde{u}=\tilde{u}_{\star}}.
\end{align}
Integrating again and expressing the solution in terms of the physical
field renders
\begin{align}
\tilde{\phi}_{00}= \frac{1}{\tilde{\rho}}(F_{00}(\tilde{v})
+G_{00}(\tilde{u})),
\end{align}
where $G_{00} = \phi_{00}|_{\tilde{v}=\tilde{v}_{\star}}$ and $F_{00}=
\int_{\tilde{v}_{\star}}^{\tilde{v}}f_{00}(s)ds$. 
Therefore,
one can consider equation \eqref{eq:high-order-asymptotic-system-good}
as the higher order asymptotic system
which is sensitive to all (higher-order) $i^0$-cylinder-logs.
To see that this is the case, it is sufficient
to compute the associated  $f_{\ell m}$ implied by the exact solution
obtained through the conformal method.
A calculation using Proposition \ref{Prop:NPtoFgauge}
and Lemma \ref{Lemma:Sol_Jacobi_and_Logs}
with $\ell=m=0$, gives
\begin{align}\label{eq:f00}
  f_{00}= \frac{2^2D_{000}}{\tilde{v}} - \sum_{p=1}^{\infty}
  \frac{ 2^{2-p}}{(p-1)!}\frac{A_{p;0,0}}{\tilde{v}^{p+1}}.
\end{align}
Therefore, the condition $f_{00}(\tilde{v})=\mathcal{O}(\tilde{v}^{-2})$
---compare with equation \eqref{eq:leadingnolog}--- implies that $D_{000}=0$.
Furthermore, for general $\ell$, one has
\begin{align}\label{eq:fvAnsatz}
  f_{\ell m}(\tilde{v})= \frac{\alpha_{\ell}D_{\ell \ell m
  }}{\tilde{v}^{2\ell +1}} + \sum_{q=0}^{\infty}\beta_{q}
  \frac{A_{\ell +1 +q, \ell, m}}{\tilde{v}^{2(\ell+1)+q}},
\end{align}
for some constants $\alpha_\ell$ and $\beta_{q}$.
Consequently, $f_{\ell m}$ is given by a expansion in negative powers of
$\tilde{v}$ where the slowest decaying term is that corresponding to
$D_{ppm}$. Thus, the regularity condition $D_{\ell \ell m }=0$, is equivalent to the
requirement:
\begin{align}
f_{\ell m}(\tilde{v}) \simeq \mathcal{O}(\tilde{v}^{-2(\ell+1)}).
\end{align}
Hence, given a functional form for $f_{\ell m}(\tilde{v})$ one can obtain a 
asymptotic expansion for the physical field $\tilde{\phi}$ by
integrating equation \eqref{eq:high-order-asymptotic-system-good}.
To do so, one rewrites equation \eqref{eq:high-order-asymptotic-system-good} as
\begin{align}\label{eq:high-order-asymptsys-NP}
 (\bme^{+})^{\ell+1}\phi_{\ell m} = \tilde{\rho}^{2(\ell+1)}f_{\ell m}
  (\tilde{v}),
\end{align}
and consider the integral curves $\Upsilon(\lambda)$ of the vector field  $\bme^{+}$
for which one has
\begin{align}
\bme^{+}\phi_{\ell m}(\lambda)= \frac{d}{d\lambda}\phi_{\ell
  m}(\lambda).
\end{align}
In the case of Minkowski spacetime, these curves are
$\Upsilon(\lambda)=(\tilde{u}_{\star}, \tilde{v}(\lambda), \vartheta^A_{\star})$
where
$\frac{d\tilde{v}}{d\lambda}=
\frac{1}{4}(\tilde{v}-\tilde{u}_{\star})^{2}$.
Solving the latter equation  explicitly gives
\begin{align}
\tilde{v}=\tilde{u}_{\star} - \frac{4c_{\star}}{2+c_{\star}\lambda},
\end{align}
with $c_{\star}=\sqrt{\tilde{v}'(\lambda_{\star})}$.
Therefore,  $\lambda\simeq \tilde{v}^{-1}$ and along $\Upsilon$
one has
\begin{align}\label{eq:high-order-asymptsys-NP-alonggammaGeneral}
  \Big(\frac{d}{d\lambda}\Big)^{\ell+1}\phi_{\ell m}(\lambda) \simeq
  \lambda^{-2(\ell+1)}f_{\ell m} (\lambda),
\end{align}
Hence, once an specific form for $f_{\ell m}$ is given,
then equation
\eqref{eq:high-order-asymptsys-NP-alonggammaGeneral} can be formally integrated
$\ell+1$ times to obtain an expansion for $\phi_{\ell m}$ and in turn, for
the physical fields $\tilde{\phi}_{\ell m}=\Theta^{-1}\phi_{\ell m}$.
For the case of an $f_{\ell m}$ given by equation
\eqref{eq:fvAnsatz} one obtains, after consecutive integration
\begin{align}
  \tilde{\phi}_{\ell m}(\lambda) \simeq \frac{1}{\tilde{\rho}} \Bigg(
  \alpha_{\ell}D_{\ell \ell m}\lambda^\ell \ln |\lambda| +
  \sum_{q=0}^{\infty}G_{q \ell m}\lambda^{q}\Bigg),
\end{align}
where $\alpha_{\ell}$  denotes some numerical constants and
$G_{q \ell m}$ depends on the initial data constants $A_{\ell +1 +q, \ell,  m}$
appearing in the expansion for $f_{\ell m}$
along with other constants appearing at each step in the integration scheme.
To write this expression in the same format as that of 
equation \eqref{goodashflat}, it is enough to recall that
$\lambda \simeq \tilde{v}^{-1}$ and
$\tilde{\rho} \simeq \tilde{v}$ along $\Upsilon$ one gets
\begin{align}\label{eq:asympt_exp_good}
  \tilde{\phi}_{\ell m} \simeq \;\; \alpha_{\ell}D_{\ell \ell m}
  \frac{\ln |\tilde{\rho}|}{\tilde{\rho}^{\ell + 1}} +
  \sum_{q=0}^{\infty}\frac{G_{q \ell m}}{\tilde{\rho}^{q+1}}.
\end{align}
Finally observe that one recovers equation
\eqref{eq:goodexpansionwithlogs} when multiplying by the corresponding
spherical harmonic $Y_{\ell m}$ and sums up summing the contributions
to each mode.  The method based on the conservation laws
\eqref{eq:cons_laws} has the advantage of making direct contact with
the $i^0$-cylinder logs as they are controlled by coefficient $D_{\ell
  \ell m}$ in the expansions.  This is not surprising since the
$f_{\ell m}$ used for this calculation was that associated to the
solution of Lemma \ref{Lemma:Sol_Jacobi_and_Logs}. However it shows
how the harmonic decomposition used for obtaining the conservation
laws is crucial so that one can can put in correspondence the
conformal method of Lemma \ref{Lemma:Sol_Jacobi_and_Logs} and some
asymptotic expansion ---such as that based on the
conservation laws \eqref{eq:cons_laws}.  Observe that giving an
specific functional form to $f_{\ell m} (\tilde{v})$ can be interpreted
as the asymptotic-system counterpart of Ansatz
\eqref{eq:Ansatz}. Nonetheless this analogy has its limits as in the
asymptotic system method one obtains the field only along outgoing
null curves ---the solution \emph{towards} $\mathscr{I}^{+}$---- while
with the conformal method one obtains the solution in a spacetime
neighbourhood of $i^0$ and $\mathscr{I}$ ---so the behaviour of the
field \emph{along} $\mathscr{I}^{+}$ is also encoded.

It is nevertheless interesting to see the consequences of assuming a different
expansion for $f_{\ell  m}$. For instance, taking
\begin{align}
  f_{\ell m}(\tilde{v})= \frac{ (\ln \tilde{v})^n}{\tilde{v}^{2\ell
      +1}} \qquad \text{for} \qquad n \geq 1,
\end{align}
and proceeding as before one would obtain
\begin{align}\label{eq:logToPower}
\tilde{\phi}_{\ell m} \simeq \frac{1}{\tilde{\rho}^{\ell+1}}
\sum_{i=0}^{n}\alpha_i (\ln \tilde{\rho})^i.
\end{align}
Notice that in this case the logarithmic terms enter non-linearly.
Whether the expansion \eqref{eq:logToPower} can be backed up by an exact
solution obtained by modifying Ansatz \eqref{eq:Ansatz} ---hence
considering a larger class of initial data--- and solving the
resulting equations exactly will be left future work.  Also it
should be point out that, although studying the behaviour of fields
propagating in flat spacetime is enough to show that the logarithmic
terms of \cite{DuaFenGasHil21, DuaFenGas22} and those of
\cite{GasVal18, Fri98a} are not the same, it would be interesting to
see if similar asymptotic expansions can be derived for asymptotically
flat spacetimes.  For generic asymptotically flat spacetimes there
will not be exact conservation laws to build the expansion upon but
plausibly one could find asymptotic conservation laws as those
described in \cite{Keh21_a} for the Schwarzschild spacetime. Using
this method or one closer to the original asymptotic system heuristics
of \cite{DuaFenGasHil21} --- but capturing the missing log terms in
the good field--- and including the ugly and bad fields will be discussed
elsewhere.

\newpage 
\section{Conclusions}
This review article gives an abridged discussion of the behaviour
close to spatial and null infinity of a system of equations called the
good-bad-ugly (GBU) model in Minkowski spacetime based on
\cite{DuaFenGasHil22b} and \cite{GasPin23}.  The good-bad-ugly system
is a simple model that mimics some of the non-linearities appearing in
a recent reformulation of the Einstein field equations in generalised
harmonic gauge well adapted to the hyperboloidal initial value problem
using the dual-foliation formalism  for numerical relativity (HypDF)
---see \cite{HilHarBug16, GasGauHil19,
  DuaFenGasHil22a}. In the hyperboloidal approach, the initial
hypersurface on which initial data is given
does not end at $i^0$ but rather at a cut of $\mathscr{I}$.  The use
of coordinates adapted to these hyperboloids brings $\mathscr{I}$ to
a finite coordinate distance but makes the Einstein field equations
formally singular at $\mathscr{I}$.  Nonetheless, as discussed in
\cite{HilHarBug16} the formally singular coefficients are multiplied
by derivatives of the metric that are expected to decay
fast enough towards null infinity. The main instrument that has been
used in the HypDF program to predict the decay the gravitational field
is a heuristic tool called the asymptotic-system. In \cite{DuaFenGasHil21} a
generalisation of the asymptotic system heuristics of \cite{GasHil18}
---see also \cite{LinRod03}---
were applied to the GBU model, first, in
Minkowski spacetime, and then, in asymptotically flat backgrounds.
Later, in \cite{DuaFenGas22} this heuristic method was applied to the Einstein
field equations in generalised harmonic gauge obtaining a
polyhomogeneous peeling result for the Weyl tensor. This result looks
similar to the polyhomogeneous result of \cite{GasVal18} derived using
conformal methods and exploiting the construction of Friedrich's
cylinder at $i^0$ in asymptotically flat spacetimes.
Then, to clarify if the logarithmic terms reported
in \cite{DuaFenGas22} and \cite{GasVal18} are the same, in
\cite{DuaFenGasHil22b} the formerly
referred conformal method was applied to GBU model in the simplest
scenario: Minkowski spacetime.  The conclusion of the analysis of
\cite{DuaFenGasHil22b} is that, the logarithmic terms reported in
\cite{DuaFenGas22} and those of \cite{GasVal18} cannot be the same.
This is emphatically evidenced by the fact that the expansion for the
good field (in flat spacetime) reported in \cite{DuaFenGasHil21} does
not contain logarithmic terms while the analysis of
\cite{DuaFenGasHil22b} (and revisited here), shows that it does.
Hence, the higher order asymptotic expansion as currently presented in
\cite{DuaFenGasHil21} missed the the logarithmic terms arising from
the degeneration of the conformal structure at $i^0$. Moreover, as
shown in Corollaries \ref{coro:logfree_data_physgood} and
\ref{coro:example_pure_log_data}, although by fine-tuning the initial
data, the good field extends analytically through the critical sets
(log-free expansion), the initial data considered in
\cite{DuaFenGasHil21} does not preclude this possibility.  Hence, the
asymptotic expansions as presented in \cite{DuaFenGasHil21} are
insensitive to the appearance of such logarithmic terms.  Nonetheless,
more recently, a possible avenue to obtain a heuristic expansion that
does not make use of conformal methods but still captures the missing
$i^0$-cylinder logarithmic terms (in flat spacetime) was given in
\cite{GasPin23}.  The alternative expansion method of \cite{GasPin23}
departs from that of \cite{DuaFenGasHil21} as it exploits the
existence of conservation laws related to the Newman-Penrose constants
---see \cite{GasPin23, Keh21_a}.  Nevertheless, it is possible that the
method of \cite{DuaFenGasHil21} can be refined without relying on a
set of (asymptotic) conservation laws and still capture the missing
logarithmic terms for the good field. If such approach is indeed
viable, it will be presented elsewhere.  As in the hyperboloidal
initial value problem the initial hypersurface does not end at $i^0$
but rather at $\mathscr{I}$ one could be led to think that the
\emph{problem at $i^0$} has no impact on the hyperboloidal
problem. However, the loss of regularity at the critical sets extends
to the rest of the conformal boundary and hence these logarithmic
terms would appear at the level of initial data on the asymptotic end
of the initial hyperboloid.


\subsection*{Acknowledgements}
 E. Gasper\'in holds a FCT (Portugal) investigator grant
2020.03845.CEECIND. Scientific discussions with
D. Hilditch, J. Feng, M. Duarte and R. Pinto are acknowledged.



\end{document}